# Pilot Reuse Factor with Large Scale Fading Precoding for Massive MIMO

**[1]Tedros Salih, [2]Elijah Mwangi, [3]Kibet Langat**

[1]Department of Electrical Engineering, Pan African University, Po Box 62000 Nairobi, Kenya

[2]School of Engineering, University of Nairobi, Po Box 30197 Nairobi, Kenya

[3]Department of Telecommunication and Information Engineering, Jomo Kenyatta University of Agriculture and Technology, Po Box 62000, Nairobi Kenya

{[1]tedysal2001@gmail.com; [2]elijah.mwangi@uonbi.ac.ke; [3]kibetlp@jkuat.ac.ke }

**Abstract**

The fundamental limitation of massive MIMO technology is pilot contamination effect. This effect occurs during uplink training when terminals use the same orthogonal signals. In this paper, a pilot reuse factor with large scale fading precoding is proposed to mitigate the pilot contamination effect. The pilot reuse factor is designed to assign unique orthogonal signals to the adjacent cells. These unique orthogonal signals are reused only within the cell and hence, intra-pilot contamination is the only concern. Large scale fading precoding is then used to mitigate the intra-pilot contamination effect. The average achievable sum rate is computed for different pilot reuse factors. Experimental results through MATLAB simulation show that a higher pilot reuse factor gives better average achievable sum rates.

## 1. INTRODUCTION

Massive MIMO technology uses a large number of base station antennas to communicate with several single-antenna users at the same time and at the same frequency [1-3].

The channel state information of massive MIMO is estimated at the base station using an uplink training based on Time Division Duplex (TDD). In the uplink training, multiple users send orthogonal signals which are known at the base station. However, in TDD operation the coherence time is very small and unique orthogonal signals for all users in all cells cannot be realized. This means that orthogonal signals will be reused within a cell or in adjacent cells. This leads to channel estimation error also referred to as pilot contamination [4] [5].

A large scale fading precoding in [6] [7] is proposed to mitigate pilot contamination that is caused by reusing orthogonal signals in adjacent cells. It is based on the assumption that the signals from all terminals in all cells are accessible at each Base Station (BS) and that slow-fading coefficients are accessible to all the BS or alternatively to a network hub. Instead of mitigating interference caused by pilot contamination in estimating of the channel, each BS uses the pilot contamination for transmitting information to all terminals. This mitigation technique gives complication of the network since it uses cooperation between cells. A large scale fading precoding with non-cooperation cell (LSFP-NCC) has been proposed in [8] to avoid cooperation of cells. In this paper, a general design of LSFP-NCC is proposed by designing a pilot reuse factor. The advantage of large scale fading precoding is used with pilot reuse factor to mitigate the pilot contamination effect. The proposed pilot reuse factor allows unique orthogonal signals to be assigned to adjacent cells. This mitigates the inter-pilot contamination effect. Since the assigned unique orthogonal signals within the cell are not enough for all users within the cell and those orthogonal signals are reused within the cell. Reusing orthogonal signals within the cell causes intra-pilot contamination. A large scale fading precoding is therefore applied within the cell to mitigate the intra-pilot contamination effect.

The rest of the paper is organized as follows. The pilot reuse factor and system model is presented and discussed in section 2 and section 3 respectively. The analysis of pilot contamination is presented in section 4, and in section 5 a discussion of the large scale fading precoding is discussed. In section 6, a derivation of the achievable rate with a finite number of base station antennas is given. In section 7, the MATLAB simulation results are presented. Finally, section 8 gives a conclusion and suggestions for future extension of the work.

## 2. PILOT REUSE FACTOR

A pilot reuse factor is designed to have each cell within a cluster unique orthogonal signals. This enables the mitigation of inter-pilot contamination effect. Even though the pilot contamination effect from adjacent cells are mitigated, the intra-pilot contamination effect still exist. Mitigation of intra-pilot contamination will be considered in Section 5. The pilot reuse factor is calculated in a similar manner to the cellular frequency reuse factor [9] given as:

$$N = i^2 + ij + j^2 \quad \textbf{(1)}$$

Where $i$ and $j$ are any non-negative integers and with $N > 1$.

From eq. (1) the following pilot reuse factor values are possible: $\{3,4,7,9,12 \ldots\}$. Fig.1 shows a hexagonal geometric arrangement of cells with a pilot reuse factor $N = 7$. Each cell within a cluster is assigned unique orthogonal signals: $PF_n = \{PF_1, PF_2, \ldots PF_7\}$. The number of unique orthogonal signals assigned to each cell within cluster is $PF_n$.

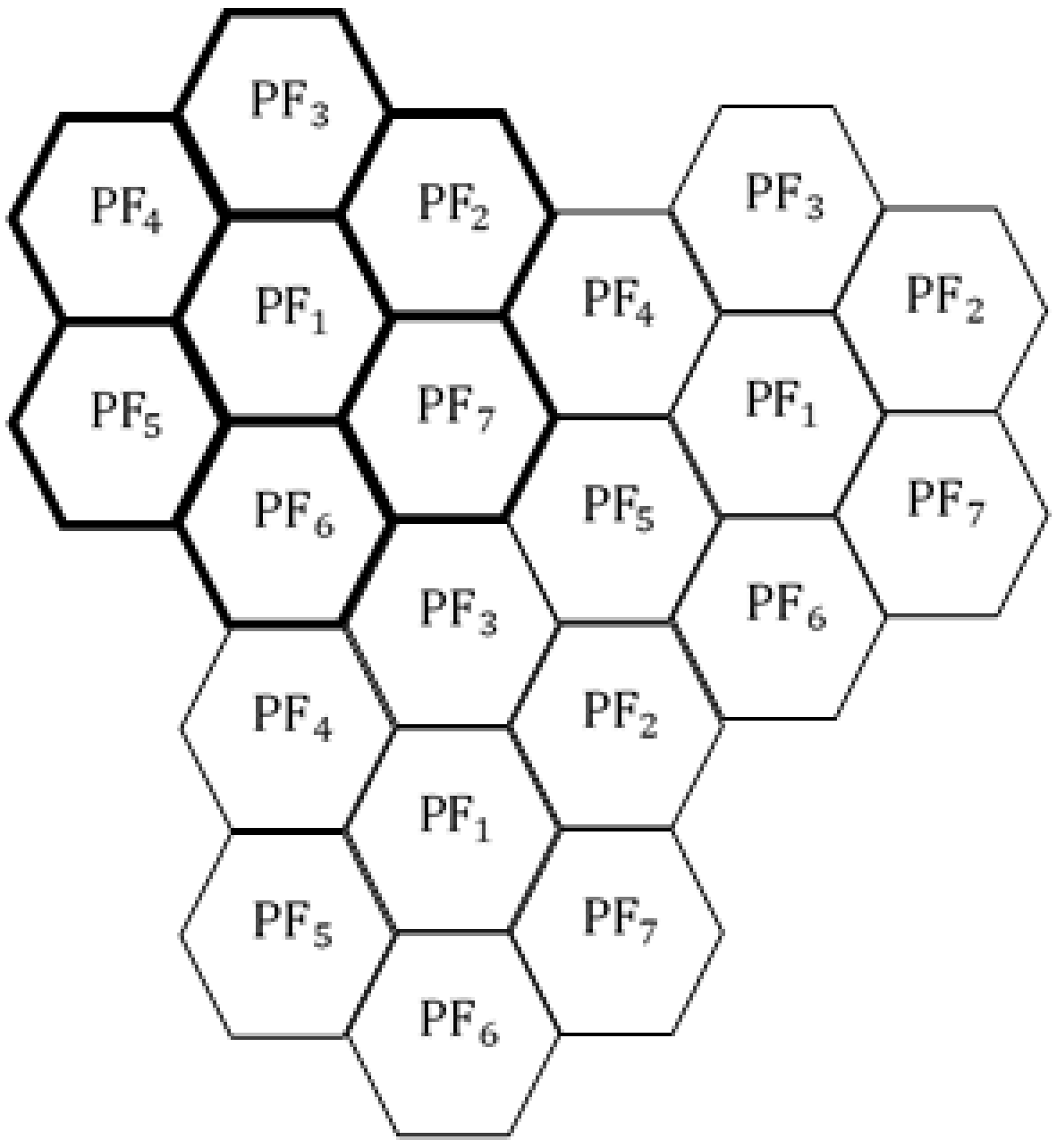

Fig. 1. Cellular hexagonal cell with $N = 7$

## 3. SYSTEM MODEL

Consider a cellular network with $L$ hexagonal cells, each consisting of one BS and $K$ single antenna users. In each cell the BS consisting of $W$ distributed sub-array antennas and each sub-array has $M$ antennas, as shown in Fig. 2.

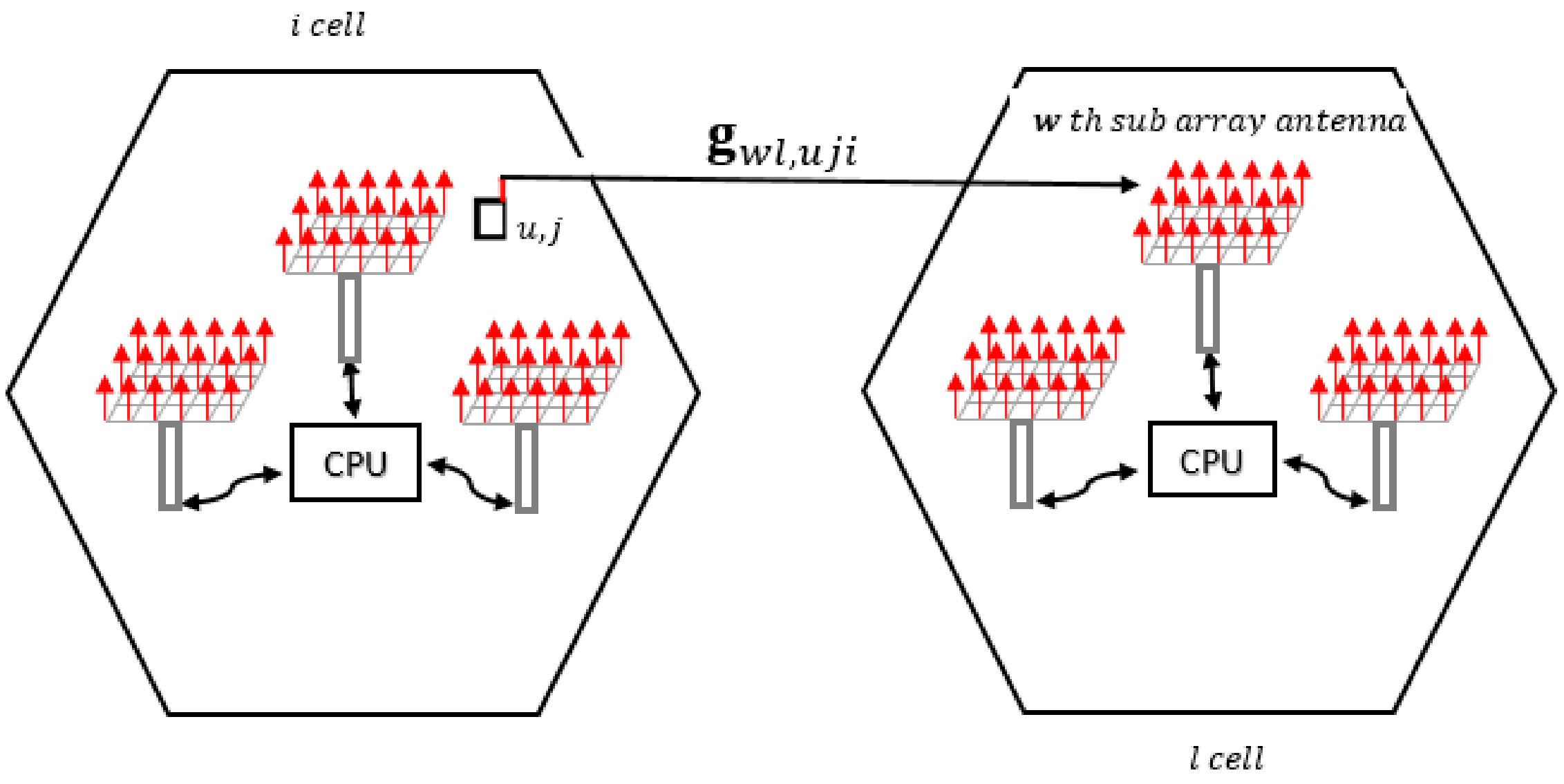

Fig. 2. Multi-cell Massive MIMO [8]

The channel for massive MIMO modeled in [4] [6] [7] [10] [11] is a combination of a small scale fading and large scale fading. The channel coefficient from $i$-th cell of $(u,j)$ user to the $l$-th cell of $w$-th sub array of antenna is given by:

$$\mathbf{g}_{wl,uji} = \boldsymbol{h}_{wl,uji}\sqrt{\beta_{wl,uji}} \qquad \textbf{(2)}$$

$$\boldsymbol{h}_{wl,uji} = (h_{1wl,uji}, h_{2wl,uji}, \dots, h_{Mwl,uji})^T \in \mathbb{C}^{M\,x\,1}$$

$$\mathbf{g}_{wl,uji} = (g_{1wl,uji}, g_{2wl,uji}, \dots, g_{Mwl,uji})^T \in \mathbb{C}^{M\,x\,1}$$

Where $\beta_{wl,uji}$ is a large scale fading and $\boldsymbol{h}_{wl,uji}$ is small scale fading with distribution of $\mathcal{CN}(0, \mathbf{I}_{\mathrm{M}})$ .

It is assumed that the $W$ sub-arrays of the BS are separated by some distance so that the slow fading differs from one sub-array to another. Since the distance between the sub-array antennas is significantly less than the distance from the sub-array to the $(u,j)$ user, the slow fading can be assumed to be independent of the $M$ antennas of the sub-array.

Each $w$-th sub-array of the BS estimates the channel using uplink training. The uplink training has the following procedure

- Using a pilot reuse factor each cell within a cluster is assigned unique $U = \frac{K}{N}$ orthogonal signals and each orthogonal signal will have a length of $\tau = K$
- Since $K > U$, all $U$ orthogonal signals are reused within the cell $J$ times
- $J$,$N$, and, $W$ are assumed to be equal.

The $U$ orthogonal signals $\boldsymbol{PF} = \{\phi_{1,l}, \phi_{2,l} \ldots \phi_{U,l}\}$ have the following property.

- Within cell $\phi_{n,l}\phi_{r,l}^{\dagger} = \delta_{nr}$
- Adjacent cells $\phi_{n,l}\phi_{r,i}^{\dagger} = 0$ .

The received signal at of $w$-th sub-array of the BS from all cell of users given by:

$$\mathbf{Y}_{l,w} = \sum_{i=1}^{\mathrm{L}} \sum_{j=1}^{\mathrm{K}} \sum_{u=1}^{\mathrm{U}} \sqrt{\tau \rho_r} \mathbf{g}_{wl,uji} \phi_{u,l} + \mathbf{n}_{l,w} \qquad (\mathbf{3})$$

Where $\mathbf{n}_{l,w} \epsilon\, \mathbb{C}^{M \times U}$ is the additive noise with distribution $\mathbf{n}_{l,w} \sim \mathcal{CN}(0,1)$ .

The $w$-th sub-array of BS estimates the channel $\mathbf{g}_{wl,uji}$ from eq. ($\mathbf{3}$)

$$\hat{\mathrm{g}}_{wl,uwl} = \mathrm{Y}_{l,w} \frac{\phi_{u,l}^{\dagger}}{\sqrt{\tau \rho_r}} = \sum_{j=1}^{\mathrm{K}} \mathrm{g}_{wl,ujl} + \frac{\phi_{u,l}^{\dagger}}{\sqrt{\tau \rho_r}} \mathbf{n}_{l,w} \qquad (\mathbf{4})$$

The $w$-th sub-array of BS pre-code the desired signal $s_{uwl}$ using the conjugated beamforming precoder $T_{wu}$ [6] [7] [10] [11]. The precoder $T_{wu}$ is given by:

$$T_{wu} = \frac{\hat{\mathbf{g}}_{wl,uwl}^{\dagger}}{\left\| \hat{\mathbf{g}}_{wl,uwl} \right\|} \qquad (\mathbf{5})$$

The $w$-th sub-array of BS sends the transmitted signal vector $T_w$ to intended user $(u, j)$. The vector $T_w$ is given by:

$$T_w = T_{wu}s_{uwl} = T_{w1}s_{1wl} + T_{w2}s_{2wl} + \cdots + T_{wU}s_{Uwl} \quad \textbf{(6)}$$

The sub-group user $(u, j)$ receives the signal:

$$x_{ujl} = \sum_{i=1}^{\mathrm{L}} \sum_{w=1}^{\mathrm{L}} \mathbf{g}_{wl,uji} T_w + \mathrm{w}_{ujl} \ , \mathrm{w}_{ujl} \ \epsilon \ \mathcal{CN}(0, \sigma_{ujl}^2) \quad (7)$$

$$x_{ujl} = \sum_{i=1}^{\mathrm{L}} \sum_{w=1}^{\mathrm{L}} \sum_{q=1}^{\mathrm{U}} \mathbf{g}_{wl,uji} T_{wq} s_{qwl} + \mathrm{w}_{ujl} \quad \textbf{(8)}$$

## 4. ANALYSIS OF THE PILOT CONTAMINATION EFFECT

In this section, the effect of pilot contamination with the proposed uplink training scheme in an asymptotic regime when $M$ tends to infinity will be investigated.

*Lemma 1.* Let $\boldsymbol{y}, \boldsymbol{z}$ $\mathbb{C}^{M\,x\,1}$ be two independent vectors with distribution $\mathcal{CN}(0, c\mathbf{I})$.

$$\lim_{M\to\infty} \frac{\boldsymbol{y}^{\dagger}\boldsymbol{z}}{M} = 0 \ \text{ and } \lim_{M\to\infty} \frac{\boldsymbol{y}^{\dagger}\boldsymbol{y}}{M} = c \quad \textbf{(9)}$$

Using the *Lemma 1*. The following property can be observed:

$$\lim_{M\to\infty} \frac{\mathbf{g}_{wl,uji}\hat{\mathbf{g}}_{wl,qwl}^{\dagger}}{M} = \begin{cases} \beta_{wl,ujl} & for\ \mathrm{q} = u\ , i = l \\ 0 & for\ \mathrm{q} \neq u \end{cases} \quad \textbf{(10)}$$

$$\lim_{M\to\infty} \frac{\left\|\hat{\boldsymbol{g}}_{wl,qwl}^{\dagger}\right\|}{\sqrt{\boldsymbol{M}}} = \boldsymbol{\mathcal{B}}_{ujl} = \left(\sum_{i=1}^{L} \boldsymbol{\beta}_{wl,uji} + \mathbf{1}\right)^{\frac{1}{2}} \quad \textbf{(11)}$$

Where $\beta'_{wl,uji} = \frac{\beta_{wl,ujl}}{\mathcal{B}_{ujl}}$

When the number of antenna approaches infinity eq.(**8**) can be written as:

$$\lim_{M\to\infty} \frac{\mathbf{g}_{wl,uji}T_{wq}}{\sqrt{M}} = \lim_{M\to\infty} \left( \frac{\mathbf{g}_{wl,uji}\hat{\mathbf{g}}^{\dagger}_{wl,qwl}}{M} \frac{1}{\frac{\left\|\hat{\mathbf{g}}^{\dagger}_{wl,qwl}\right\|}{\sqrt{M}}} \right) \quad (12)$$

$$\lim_{M\to\infty} \frac{x_{ujl}}{\sqrt{M}} = X_{ujl} = \sum_{w=1}^{\mathrm{L}} \beta'_{wl,ujl} s_{uwl} \quad (13)$$

From eq. (**13**), two observations can be made: The first is that the signal that comes from adjacent cell vanishes when $M$ approaches infinity. The second is that the $(u, j)$-th user at $l$-th cell not only receives the intended signal from the expected sub-array BS of the $l$-th cell, but it also receives signals from other sub-array BS of the $l$-th cell. This causes signal interference to occur. The main reason for this interference is the pilot contamination effect.

## 5. LARGE SCALE FADING PRECODING

To mitigate the pilot contamination effect the $w$-th sub-array can be use additional pre-coder of some unknown variable set $\{C_1,\ C_2 \ldots C_L\}$ so as to get the received signal eq. (**13**) as:

$$X_{ujl} = \sum_{w=1}^{\mathrm{L}} \beta'_{wl,ujl} v_{uwl} = s_{ujl} \quad (14)$$

Where;

$$v_{uwl} = \mathrm{C_w} s_{uwl}$$

Eq. (**14**) can be re-written as:

$$\begin{aligned}
X_{u1l} &= \beta'_{1l,u1l} v_{u1l} + \beta'_{2l,u1l} v_{u2l} + \cdots + \beta'_{Ll,u1l} v_{uLl} = s_{u1l} \\
X_{u2l} &= \beta'_{1l,u2l} v_{u1l} + \beta'_{2l,u2l} v_{u2l} + \cdots + \beta'_{Ll,u2l} v_{uLl} = s_{u2l} \\
&\vdots \\
X_{uLl} &= \beta'_{1l,uLl} v_{u1l} + \beta'_{2l,uLl} v_{u2l} + \cdots + \beta'_{Ll,uLl} v_{uLl} = s_{u2l}
\end{aligned} \quad (15)$$

Then eq. (**15**) can be given in matrix form as:

$$\mathbf{BV} = \mathbf{S} \qquad \textbf{(16)}$$

Where;

$$\boldsymbol{B} = \begin{pmatrix} \beta'_{1l,u1l} & \beta'_{2l,u1l} & \cdots & \beta'_{Ll,u1l} \\ \beta'_{1l,u2l} & \beta'_{2l,u2l} & \cdots & \beta'_{Ll,u2l} \\ \vdots & \vdots & \ddots & \vdots \\ \beta'_{1l,uLl} & \beta'_{2l,uLl} & \cdots & \beta'_{Ll,uLl} \end{pmatrix}, \quad \boldsymbol{V} = \begin{pmatrix} v_{u1l} \\ v_{u2l} \\ \vdots \\ v_{uLl} \end{pmatrix}, \quad \boldsymbol{S} = \begin{pmatrix} s_{u1l} \\ s_{u2l} \\ \vdots \\ s_{uLl} \end{pmatrix}$$

Eq. (**16**) can also be written as:

$$\mathbf{V} = \mathbf{AS} \qquad \textbf{(17)}$$

Where;

$$\boldsymbol{A} = \begin{pmatrix} a_{1l,u1l} & a_{2l,u1l} & \cdots & a_{Ll,u1l} \\ a_{1l,u2l} & a_{2l,u2l} & \cdots & a_{Ll,u2l} \\ \vdots & \vdots & \ddots & \vdots \\ a_{1l,uLl} & a_{2l,uLl} & \cdots & a_{Ll,uLl} \end{pmatrix}$$

The matrix $\boldsymbol{A}$ is inverse of the matrix **B** is given by:

$$\boldsymbol{A} = \boldsymbol{B}^{-1} \qquad \textbf{(18)}$$

The $w$-th sub-array of BS pre-code the desired signal $v_{uwl}$ using a conjugated beamforming $T_{wu}$ and sends the vector $T_w$ to the intended user $(u, j)$ . This precoding scheme is termed large scale fading precoding. The vector $T_w$ is given by:

$$T_w = \sum_{u=1}^{\mathrm{U}} T_{wu} v_{uwl} \qquad \textbf{(19)}$$

The sub-group user $(u, j)$ receives the signal

$$x_{ujl} = \sum_{i=1}^{L} \sum_{w=1}^{L} \sum_{q=1}^{U} \boldsymbol{g}_{wl,uji} T_{wq} s_{qwl} + \boldsymbol{w}_{ujl} \quad \textbf{(19)}$$

$$\lim_{M \to \infty} \frac{x_{ujl}}{\sqrt{M}} = X_{ujl} = \sum_{w=1}^{\mathrm{L}} \beta'_{wl,ujl} v_{uwl} = s_{ujl} \quad \textbf{(20)}$$

From eq. (**20**) it can be concluded that the proposed pilot reuse factor with the large scale fading precoding fully cancels the pilot contamination effect in an asymptotic regime.

## 6. FINITE SUB-ARRAY ANTENNAS EFFECT

In this section, the achievable rate of the received signal when a finite number of antennas are used is derived. Considering MMSE estimator, the propagation channel can be estimated from the received signal of eq.(**3**) to be:

$$\hat{\mathbf{g}}_{wl,uwl} = \mathbf{Y}_{l,w}\left(\phi_{u,l}^{\dagger} \varphi'_{wl,uwl}\right) = \varphi'_{wl,uwl} \sqrt{\tau \rho_t} \sum_{j=1}^{L} \mathbf{g}_{wl,ujl} + \boldsymbol{n}'_{l,w} \quad \textbf{(21)}$$

Where;

$$\varphi'_{wl,uwl} = \frac{\sqrt{\tau \rho_t} \beta_{wl,uwl}}{\zeta^2_{wl,uwl}}, \qquad \zeta^2_{wl,uwl} = 1 + \rho_t \tau \sum_{z=1}^{L} \beta_{wl,uzl} \quad \textbf{(22)}$$

The vectors $\mathbf{g}_{wl,uji}$ and $\hat{\mathbf{g}}_{wl,uwl}$ have the following distributions**;**

$$\mathbf{g}_{wl,uji} \sim \mathcal{CN}\left(0, \beta_{lwl,uji} \boldsymbol{I_M}\right) \quad \textbf{(23)}$$

$$\hat{\mathbf{g}}_{wl,uwl} \sim \mathcal{CN}\left(0, \left(\frac{\sqrt{\tau \rho_t} \beta_{wl,uwl}}{\zeta_{wl,uwl}}\right)^2 \boldsymbol{I_M}\right) \quad \textbf{(24)}$$

The sub-group user $(u, j)$receives the signal:

$$x_{ujl} = \sqrt{\rho_b} \sum_{i=1}^{\mathrm{L}} \sum_{w=1}^{\mathrm{L}} \sum_{q=1}^{\mathrm{U}} \mathbf{g}_{wl,uji} \frac{\hat{\mathbf{g}}_{wl,qwl}^{\dagger}}{\zeta_{wl,qwl}} v_{qwi} + \mathrm{w}_{ujl} \quad \textbf{(25)}$$

The conjugated beamforming is assumed to be: $T_{wq} = \frac{\hat{\mathbf{g}}_{wl,qwl}^{\dagger}}{\zeta_{wl,qwl}}$

After some manipulation, eq.(**25**) can be written as:

$$x_{ujl} = \mathrm{v}_1 + \mathrm{v}_2 + \mathrm{v}_3 + \mathrm{v}_4 + \mathrm{v}_5 + \mathrm{v}_6 \quad \textbf{(26)}$$

Where;

$$\mathrm{v}_1 = s_{ujl}\sqrt{\rho_b} \sum_{w=1}^{\mathrm{L}} \boldsymbol{a}_{wl,ujl} \mathrm{E}\left[\mathbf{g}_{wl,ujl} \frac{\hat{\mathbf{g}}_{wl,uwl}^{\dagger}}{\zeta_{wl,uwl}^2}\right]$$

$$\mathrm{v}_2 = s_{ujl}\sqrt{\rho_b} \sum_{w=1}^{\mathrm{L}} \boldsymbol{a}_{wl,ujl} \left(\mathbf{g}_{wl,ujl} \frac{\hat{\mathbf{g}}_{wl,uwl}^{\dagger}}{\zeta_{wl,uwl}^2} - \mathrm{E}\left[\mathbf{g}_{wl,ujl} \frac{\hat{\mathbf{g}}_{wl,uwl}^{\dagger}}{\zeta_{wl,uwl}^2}\right]\right)$$

$$\mathrm{v}_3 = \sqrt{\rho_b} \sum_{w=1}^{\mathrm{L}} \mathbf{g}_{wl,ujl} \frac{\hat{\mathbf{g}}_{wl,uwl}^{\dagger}}{\zeta_{wl,uwl}^2} \sum_{\substack{p=1 \\ p \neq j}}^{\mathrm{L}} s_{upl} \boldsymbol{a}_{wl,upl}$$

$$\mathrm{v}_4 = \sqrt{\rho_b} \sum_{w=1}^{\mathrm{L}} \sum_{\substack{q=1 \\ q \neq u}}^{\mathrm{U}} \mathbf{g}_{wl,ujl} T_{wq} v_{qwl}$$

$$\mathrm{v}_5 = \sqrt{\rho_b} \sum_{\substack{i=1 \\ i \neq l}}^{\mathrm{L}} \sum_{w=1}^{\mathrm{L}} \sum_{q=1}^{\mathrm{U}} \mathbf{g}_{wl,uji} T_{wq} v_{qwi}$$

$$\mathrm{F}_6 = \mathrm{w}_{ujl}$$

The achievable rate is given by:

$$\mathrm{C} = \mathbf{log_2}\left(\mathbf{1} + \frac{\mathrm{E}[|\mathrm{v}_1|^2]}{\mathrm{E}[|\mathrm{v}_2|^2] + \mathrm{E}[|\mathrm{v}_3|^2] + \mathrm{E}[|\mathrm{v}_4|^2] + \mathrm{E}[|\mathrm{v}_5|^2] + \mathrm{E}[|\mathrm{v}_6|^2]}\right) \quad \textbf{(27)}$$

Similar to [7] it can be shown that:

$$\mathrm{E}[|\mathrm{v}_1|^2] = \rho_b M^2 \left| \sum_{w=1}^{\mathrm{L}} \frac{\rho_t \tau \beta_{wl,ujl} \beta_{wl,uwl}}{1 + \rho_t \tau \sum_{z=1}^{L} \beta_{wl,uzl}} \frac{\boldsymbol{a}_{wl,ujl}}{\zeta_{wl,uwl}} \right|^2 \quad \textbf{(28)}$$

$$E[|v_2|^2] = \rho_b M \sum_{w=1}^{L} \left|\frac{\boldsymbol{a}_{wl,ujl}}{\zeta_{wl,uwl}}\right|^2 \frac{\rho_t \tau \beta_{wl,ujl} \beta_{wl,uwl}^2}{1 + \rho_t \tau \sum_{z=1}^{L} \beta_{wl,uzl}} \quad \textbf{(29)}$$

$$E[|v_3|^2] = \rho_b M \sum_{w=1}^{L} \sum_{\substack{\boldsymbol{p=1} \\ \boldsymbol{p \neq j}}}^{L} \left|\frac{\boldsymbol{a}_{wl,upl}}{\zeta_{wl,uwl}}\right|^2 \frac{\rho_t \tau \beta_{wl,ujl} \beta_{wl,uwl}^2}{1 + \rho_t \tau \sum_{z=1}^{L} \beta_{wl,uzl}}$$

$$+ \rho_b M^2 \sum_{\substack{\boldsymbol{p=1} \\ \boldsymbol{p \neq j}}}^{L} \left| \sum_{w=1}^{L} \frac{\rho_t \tau \beta_{wl,ujl} \beta_{wl,uwl}}{1 + \rho_t \tau \sum_{z=1}^{L} \beta_{wl,uzl}} \frac{\boldsymbol{a}_{wl,upl}}{\zeta_{wl,uwl}} \right|^2 \quad \textbf{(30)}$$

$$E[|v_4|^2] = \rho_b M \sum_{w=1}^{L} \sum_{\substack{q=1 \\ q \neq u}}^{U} \frac{\rho_t \tau \beta_{lwl,ujl} \beta_{wl,qwl}^2}{1 + \rho_t \tau \sum_{z=1}^{L} \beta_{wl,qzl}} \left|\frac{\boldsymbol{v}_{qwl}}{\zeta_{wl,qwl}}\right|^{\mathbf{2}} \quad \textbf{(31)}$$

$$E[|v_5|^2] = \rho_b M \sum_{\substack{i=1 \\ i \neq l}}^{L} \sum_{w=1}^{L} \sum_{q=1}^{U} \frac{\rho_t \tau \beta_{lwl,uji} \beta_{wl,qwl}^2}{1 + \rho_t \tau \sum_{z=1}^{L} \beta_{wl,qzl}} \left|\frac{\boldsymbol{v}_{qwi}}{\zeta_{wl,qwi}}\right|^{\mathbf{2}} \quad \textbf{(32)}$$

$$E[|v_6|^2] = var(\mathbf{w}_{ujl}) = \sigma_{ujl}^2 . \quad \textbf{(33)}$$

Finally, the achievable rate becomes;

$$C = \mathbf{log_2}\left(\mathbf{1} + \frac{\rho_b M^2 \left| \sum_{w=1}^{L} \frac{\rho_t \tau \beta_{wl,ujl} \beta_{wl,uwl}}{1 + \rho_t \tau \sum_{z=1}^{L} \beta_{wl,uzl}} \frac{\boldsymbol{a}_{wl,ujl}}{\zeta_{wl,uwl}} \right|^2}{M^2 \psi_1 + M \psi_2 + \sigma_{ujl}^2}\right) \quad \textbf{(34)}$$

Where;

$$\psi_1 = \rho_b \sum_{\substack{\boldsymbol{p=1} \\ \boldsymbol{p \neq j}}}^{L} \left| \sum_{w=1}^{L} \frac{\rho_t \tau \beta_{wl,ujl} \beta_{wl,uwl}}{1 + \rho_t \tau \sum_{z=1}^{L} \beta_{wl,uzl}} \frac{\boldsymbol{a}_{wl,upl}}{\zeta_{wl,uwl}} \right|^2$$

$$\psi_2 = \rho_b \sum_{i=1}^{L} \sum_{w=1}^{L} \sum_{q=1}^{U} \frac{\rho_t \tau \beta_{lwl,uji} \beta_{wl,qwl}^2}{1 + \rho_t \tau \sum_{z=1}^{L} \beta_{wl,qzl}} \left|\frac{\boldsymbol{v}_{qwi}}{\zeta_{wl,qwi}}\right|^{\mathbf{2}}$$

## 7. SIMULATION RESULTS

In this section, a cellular hexagonal network with pilot reuse factor $N = 3, N = 4$, and $N = 7$ is considered. All the parameters are taken from 3GPP standard [12]. The large scale fading is modelled using a 3GPP standard of Urban Macro model given by eq. (**35**) [12].

$$10\log_{10}\beta_{wl,uji} = -139.5 - 35\log_{10}d_{wl,uji} + \varphi_{wl,uji} \qquad \textbf{(35)}$$

Where $d_{wl,uji}$ is the distance (in km) between the user and the base station and $\varphi_{wl,uji}$ is shadowing coefficient modeled as a Gaussian random variable with zero mean and variance 8dB. $d_{wl,uji}$ is the distance (in km) of all users randomly distributed near the edge of the cell. Table 1 refers to the value of parameters that has been used for simulation purpose.

Table 1 System parameters for the simulation**.**

| Parameter | Value |
|---|---|
| Shadowing coefficient $\varphi_{wl,uji}$ | $\mathcal{N}(0,8\text{dB})$ |
| cell radius $r$ | 0.75km |
| Noise variance at each receiver $\sigma^2_{ujl}$ | 92dBm |
| Average power of $w$-th sub-array $\rho_b$ | 48dBm |
| Average power at each user terminal $\rho_t$ | 23dBm |
| Bandwidth $B$ | 20MHz |

The achievable sum rate of eq. (**34**) for pilot reuse factor $N = 3, N = 4$, and $N = 7$ were obtained by MATLAB simulation.

Fig. 3 shows result of the simulation for pilot reuse factor $N = 3$ and $N = 4$ with the number of user in each cell being 48. It can be observed that pilot reuse factor $N = 4$ gives better achievable sum rate than pilot reuse factor $N = 3$. For example, for the number of antennas $M = 100$ the average achievable sum rate for pilot reuse factor $N = 3$ is $0.205 \times 10^{-2}$. The average achievable sum rate with a pilot reuse factor $N = 4$ for $M = 100$ is $0.237 \times 10^{-2}$. This is larger than the value obtained with the pilot reuse factor $N = 3$.

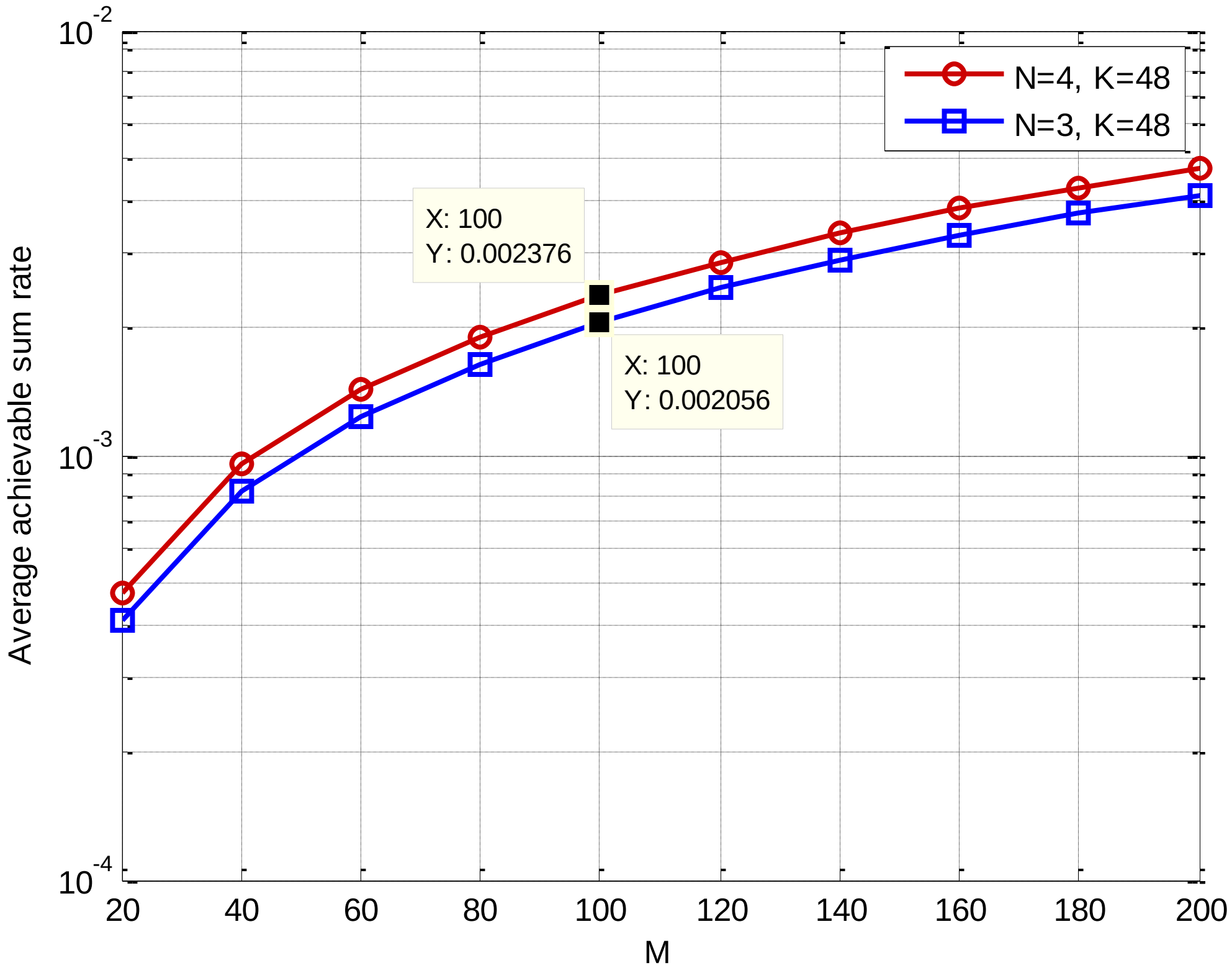

Fig. 3. The Average achievable sum rate of 48 users for pilot reuse factor $N = 4$ and $N = 3$ with different M

Fig. 4 shows result of the simulation for pilot reuse factor $N = 4$ and $N = 7$ with number of user in each cell 56. It can be observed that pilot reuse factor $N = 7$ gives better achievable sum rate than pilot reuse factor $N = 4$. For example, with the number of antennas $M = 100$ the average achievable sum rate for pilot reuse factor $N = 4$ is $.245{\times}10^{-2}$ . For pilot reuse factor $N = 7$ when $M = 100$ is $0.516{\times}10^{-2}$, which is larger than the value obtained with a pilot reuse factor $N = 4$.

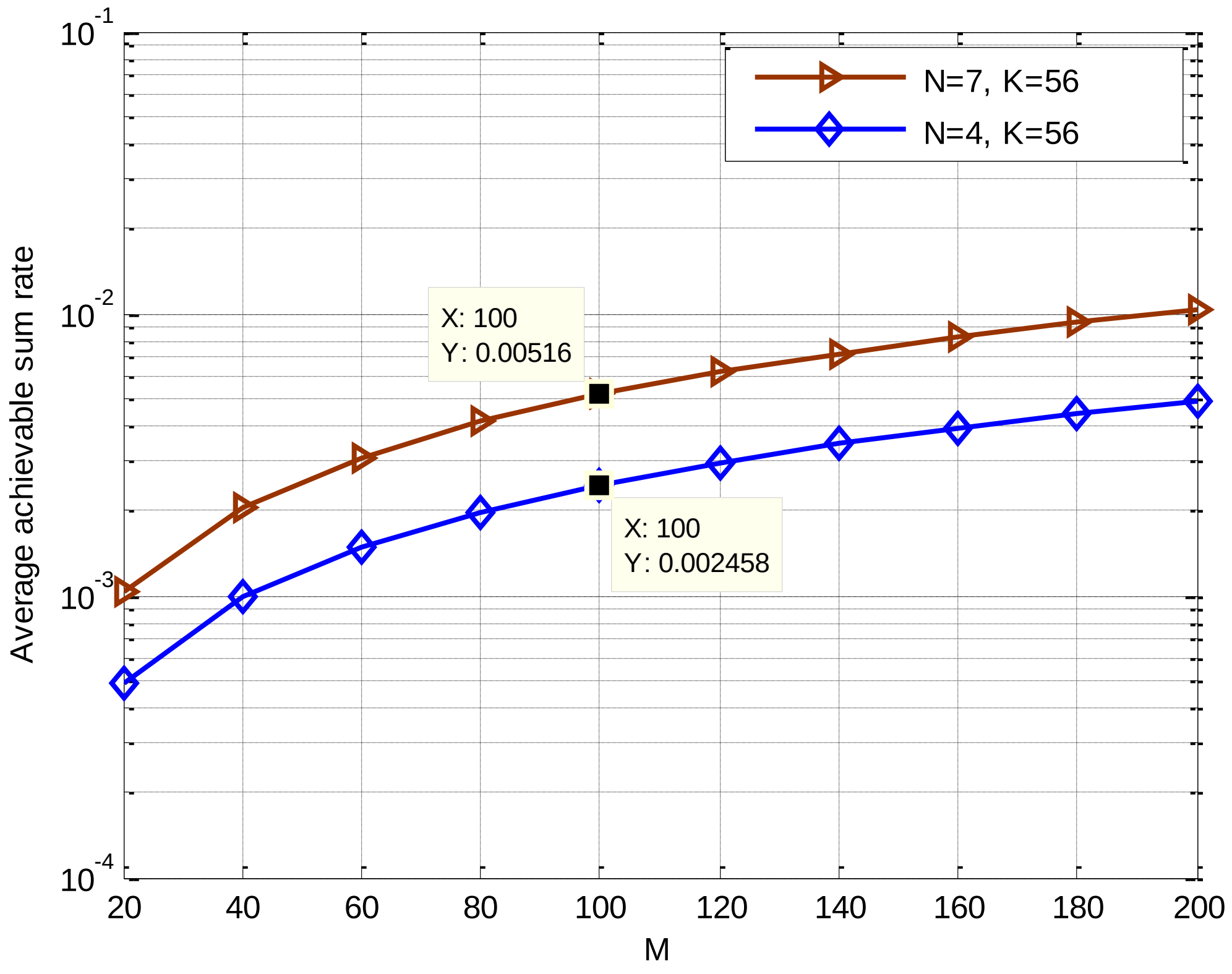

Fig. 4. The Average achievable sum rate of 56 users for pilot reuse factor $N = 7$ and $N = 4$ with different M

Fig. 5 shows a comparison of pilot reuse factor $N = 7$ with a large scale fading precoding for cooperation of cells ($N = 1$). It can be observed that the achievable sum rate improves when the pilot reuse factor is employed. For example, for the number of antennas $M = 100$ the average achievable sum rate with no the pilot reuse factor ( $N = 1$) is $0.118{\times}10^{-2}$ . For pilot reuse factor $N = 7$ when $M = 100$ is $0.36{\times}10^{-2}$ which is larger than the value obtained with no the pilot reuse factor ( $N = 1$).

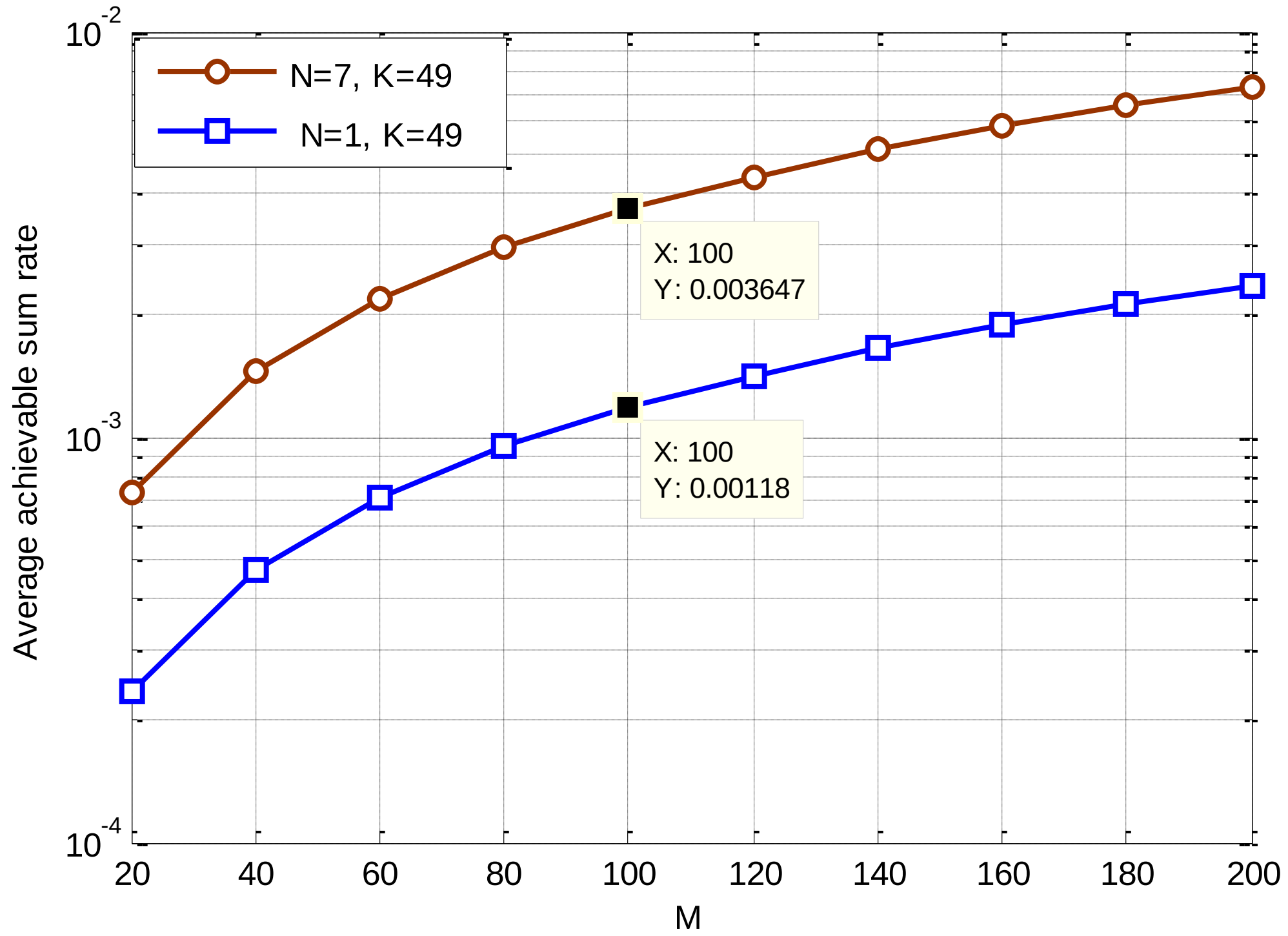

Fig. 5. The Average achievable sum rate of 49 users for $N = 7$ and $N = 1$ with different M

Fig. 6 shows a comparison of pilot reuse factor $N = 4$ with a large scale fading precoding for cooperation of cells ($N = 1$). It can be observed that pilot reuse factor $N = 4$ gives better achievable sum rate than with no the pilot reuse factor ( $N = 1$). For example, for the number of antennas $M = 100$ the average achievable sum rate for pilot reuse factor $N = 1$ is $0.112 \times 10^{-2}$ . For pilot reuse factor $N = 4$ when $M = 100$ is $0.23 \times 10^{-2}$ which is larger than the value obtained with no the pilot reuse factor ($N = 1$).

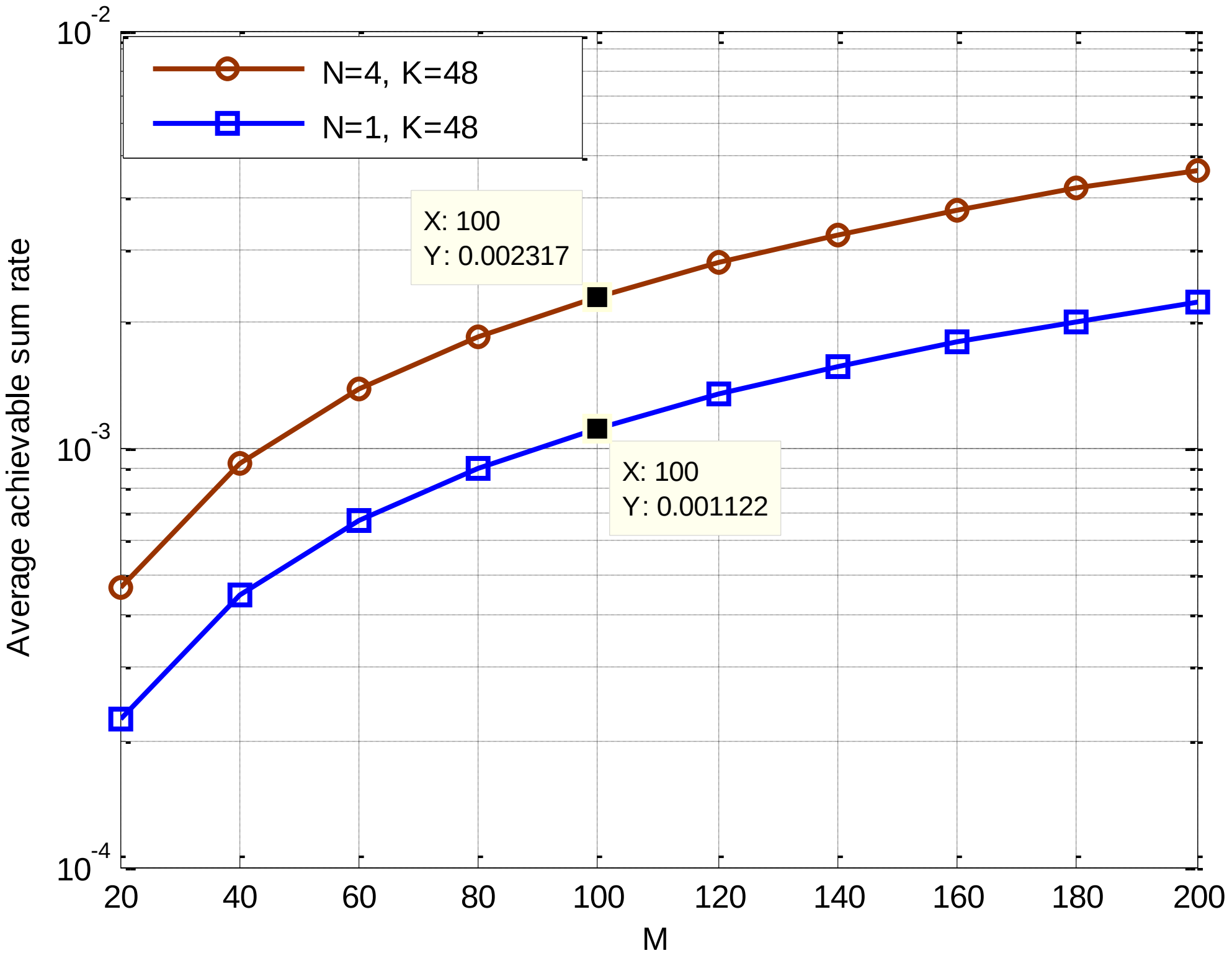

Fig. 6. The Average achievable sum rate of 48 users for $N = 4$ and $N = 1$ with different M

All the above results show that as the number of antennas increases the achievable rate also increases. It can also be observed that a higher pilot reuse factor gives better average achievable sum rates.

## 8. CONCLUSION

A pilot reuse factor with large scale fading precoding has been developed. The pilot reuse factor is designed to mitigate the effect of inter-pilot contamination. A large scale fading precoding is used to mitigate the effect of intra-pilot contamination. It has been observed that the higher pilot reuse factor gives better average achievable sum rates.

It can be noted that to obtain a better average achievable sum rate, then a higher pilot reuse factor is needed. This increases the number of sub-array base stations with a subsequent increase in installation costs and power consumption.

Further investigations needs to be carried out to determine the optimal point of pilot reuse factor so as to reduce the number of sub-array BS within the cell without any significant reduction in the average achievable sum rate.